\documentstyle[aaspp4,epsf,12pt]{article}

\newcommand{\COBE}{{\sl COBE\/}}
\newcommand{\MAP}{{\sl MAP\/}}
\lefthead{Lawrence, Scott \& White}
\righthead{What has the CMB ever done for us?}

\begin{document}

\title{What have we already learned from the CMB?}

\author{Charles R.~Lawrence${}^1$, Douglas Scott${}^2$ and Martin White${}^3$}

\vspace{0.2in}
\affil{${}^1$Jet Propulsion Laboratory,\\
169-506, 4800 Oak Grove Drive,\\
Pasadena, CA 91109\ \ U.S.A.\\
\vspace{0.1in}
${}^2$Department of Physics \& Astronomy,\\
129-2219 Main Mall, University of British Columbia,\\
Vancouver, B.C.\ \  V6T 1Z4\ \ Canada\\
\vspace{0.1in}
${}^3$Departments of Astronomy \& Physics,\\
University of Illinois at Urbana-Champaign,\\
103 Astronomy Building, 1002 West Green Street,\\
Urbana, IL 61801\ \ U.S.A.}

\authoremail{crl@jplsp2.jpl.nasa.gov, dscott@astro.ubc.ca,
white@physics.uiuc.edu}

\newpage

\begin{abstract}
\noindent
\rightskip=0pt
The COsmic Background Explorer (\COBE) satellite, and the Differential
Microwave Radiometer (DMR) experiment in particular, was
extraordinarily successful.  However, the DMR results were announced about
7 years ago, during which time a great deal more has been learned about
anisotropies in the Cosmic Microwave Background (CMB).
We assess the current state of knowledge, and discuss where we might be going.
The CMB experiments currently being designed and built, including long-duration
balloons, interferometers, and two space missions, promise to address several
fundamental cosmological issues.
We present our evaluation of what we already know, what we are beginning to
learn now, and what the future may bring.
\end{abstract}

\keywords{cosmic microwave background -- cosmology: observations
-- cosmology: theory -- large-scale structure}

\newpage

\noindent
{\sl All right. But apart from the sanitation, the medicine, education,
        wine, public order, irrigation, roads, the fresh water system, and
        public health \dots What have the Romans ever done for us?}
\vspace{4pt}

\rightline{Reg, spokesman for the People's Front of Judea\footnote{\cite{Reg}}}
\vspace{0.2in}

\section{Introduction}

\noindent
The study of the cosmic microwave background (CMB) radiation has had a long
history.
Three aspects of the CMB might be considered: its existence, its spectrum,
and its anisotropies.  By firmly establishing that the Universe expanded from
an initially hot, dense state, the {\it existence\/} of the CMB underpins our
entire cosmological framework.  It has been recognized from the beginning as
one of the pillars of the hot big bang cosmologies.
The spectrum of the CMB is the most precise blackbody spectrum in nature,
{}from which many inferences can be made.
Although this discovery is less than a decade old, its impact on models of
the early Universe been discussed extensively elsewhere, (e.g.~\cite{NorSmo}).
In this paper we would like to consider the {\it anisotropies\/} in the CMB,
the small fluctuations imprinted on the sky by the progenitors of the
large-scale structure seen in the distribution of galaxies today.

In the roughly seven years since the \COBE\ DMR team announced the first
detection of anisotropies in the CMB (\cite{Smoot}),
more than a dozen groups have
reported detections, covering the full range of frequencies and a wide range of
angular scales (see \cite{SmoSco,BenTurWhi}).  Due in large part to a dramatic
increase in detector sensitivity, mapping the CMB anisotropy has become almost
routine.  Our confidence in the results has grown as multiple observations by
the same teams over a period of years, and then later by different experiments
at different frequencies and sites, reproduced the same features on the sky and
confirmed their black body nature.

Over the same period much progress has been made in data analysis techniques
and in the theoretical interpretation of CMB data.  Better physical
understanding of the anisotropy generation has lead to faster algorithms for
its computation (e.g.~\cite{SelZal,HSWZ}) applicable to an impressively wide
range of theories.  The high precision calculations and accurate measurements
of the anisotropy have spawned numerous ideas in data analysis, with a full
likelihood analysis of mega-pixel CMB maps now within reach
(\cite{OhSpeHin}).

However, since much of the progress has been incremental, it is not always
obvious just how far we have really advanced. It therefore seems appropriate
to take stock and ask the question:\newline

\centerline{{\sl What has the CMB ever done for us?}}
\vspace{1.0in}

\section{The Lists}

\noindent
{}From a broad perspective, the main impact of CMB anisotropies has been
to shrink substantially the range of cosmological models under active
discussion.  This is not always easy to see, since the {\it number\/} of
models proposed at any time seems to be determined more by the number of
theorists working in the field than by any constraints provided by the data.
Moreover,
it sometimes seems that no class of model has been ruled out. However,
looking back a decade in the literature makes it clear that this is not true.

\begin{figure}[t]
\begin{center}
\leavevmode
\epsfxsize=16cm \epsfbox{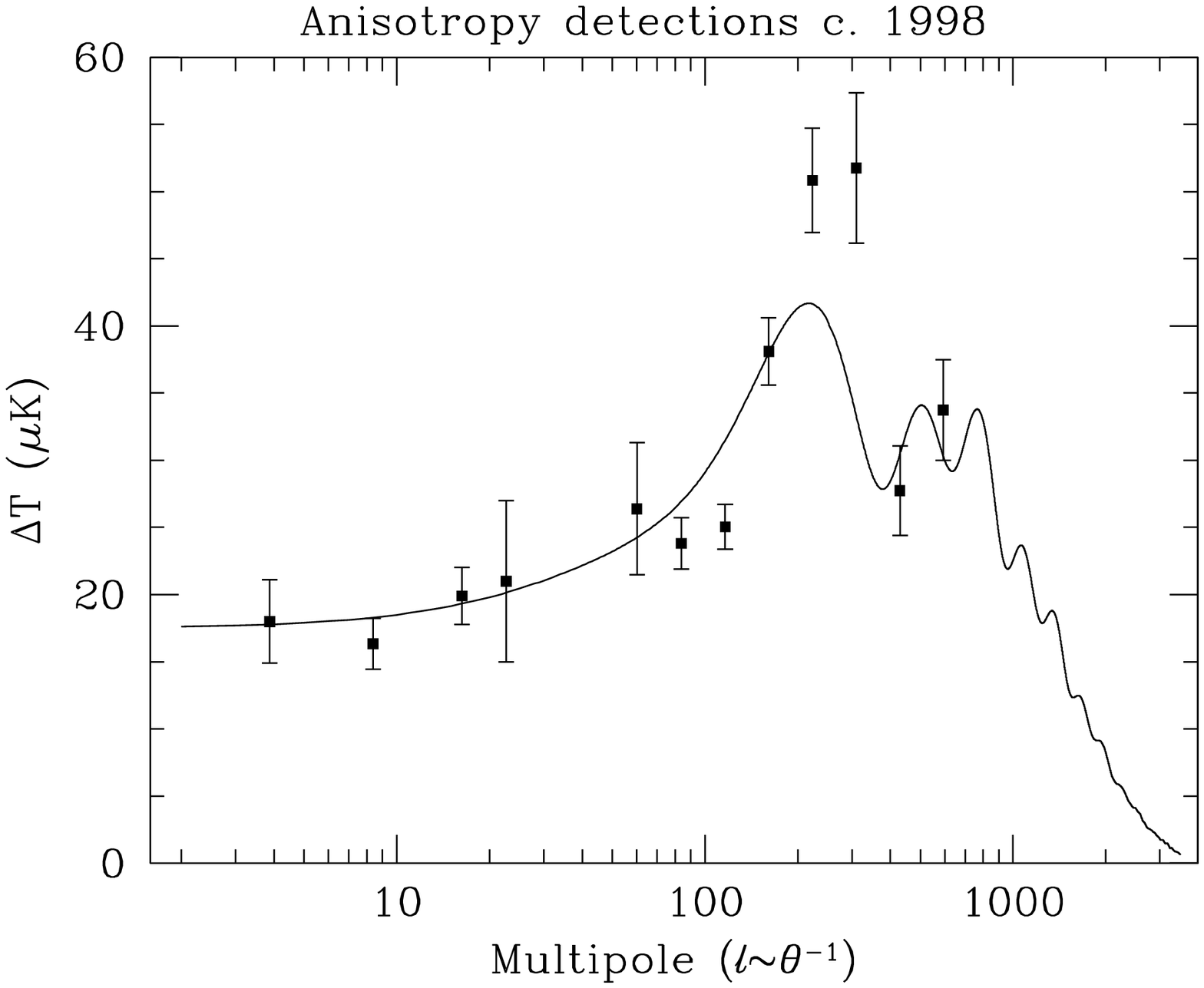}
\end{center}
\caption{The current CMB anisotropy detections, averaged in 12 bins equally
spaced in $\log\ell$ (with some bins missing, where no experimental window
functions peak).  The $y$-axis measures the rms fluctuation averaged
over the range of angular scales within the bin, the $x$-axis is the
multipole number $\ell\sim\theta^{-1}$, with $1^\circ$ near $\ell\sim 10^2$.
The solid line is the prediction of the `standard' cold dark matter model,
and is included only as an example. We note that creating plots like this
is cosmetology rather than cosmology; such binned data are qualitatively
useful, but should not be used for statistical purposes.}
\label{fig:current}
\end{figure}

\placefigure{fig:current}

Even before \COBE, the high level of isotropy of the CMB was perhaps the best
possible evidence that the large-scale properties of the Universe were well
described by the Friedman-Robertson-Walker metric. The assumption of
homogeneity
and isotropy, initially made for purely aesthetic reasons, turned out to be an
extremely good approximation to the real Universe. As the limits on anisotropy
became stronger and stronger, the number of models based on anything but the FRW
metric became fewer and fewer.

Currently popular models assume that the matter in the FRW Universe is
composed {\it mostly\/} of Cold Dark Matter (CDM), with smaller admixtures
of baryons and perhaps massive neutrinos, plus curvature and/or vacuum
components.
For these CDM-inspired models, the CMB data have been instrumental in narrowing
the range of possibilities, and most popular flavours of CDM now give
remarkably similar predictions.
Models dominated by Hot Dark Matter, already in trouble
before \COBE, are no longer discussed.
Two other classes of models, namely defects and isocurvature models, have not
been ruled out definitively, but they are now very much on the defensive
against
the weight of data.
Explosion models
(\cite{OstCow,Ikeuchi,CarIke,VisOstBer,Wandel,OstStr,WeiDekOst}),
super-conducting cosmic string models (\cite{OTW,OstTho,BorOstWei}), and
late-time phase transition models
(\cite{Wasserman,HilSchFry,PreRydSpe,FulSch,FriHilWat,JafSteFri}) have
essentially vanished.

Figure~\ref{fig:current} shows the current state of CMB measurements.
Included are all detections we are aware of that have been published or
submitted for publication in 1998.  The results have been averaged in 12 bins,
equally spaced in $\log\ell$ for clarity, and
we have omitted the upper limits on smaller angular scales, most of which
are off the right of the plot with our chosen $\ell$-axis range.  This
figure is meant to be indicative only.  More statistically rigorous approaches
exist for combining data sets (e.g.~\cite{BonJafKno}), and such methods should
certainly be used for determining precise constraints on models.
However, Fig.~\ref{fig:current} gives
approximately the correct visual impression for the combined constraining power
of today's data.

Below we list two sets of statements that we believe are supported by
the data: the first set contains `fundamental truths' about
the Universe; and the second contains statements that will be fundamental
truths if confirmed, but that for the present must be regarded more
tentatively.

\vspace{0.5in}

\noindent Here is the `A' list:
\begin{itemize}
\item[{\bf A1}]\ Gravitational instability in a dark matter dominated universe
grew today's structure
\item[{\bf A2}]\ The Universe (re)combined
\item[{\bf A3}]\ There is an excess of temperature fluctuations at roughly
the predicted angular scale
\item[{\bf A4}]\ The polarization of the CMB anisotropy is small
\end{itemize}

\vspace{0.5in}
\vfil\eject

\noindent And the `B' list:
\begin{itemize}
\item[{\bf B1}]\ Something like inflation produced adiabatic fluctuations
\item[{\bf B2}]\ The large-scale structure of space-time appears to be simple
\item[{\bf B3}]\ The gravity wave contribution to the anisotropy is not large
\item[{\bf B4}]\ There are constraints on non-standard physics at
$z\sim10^3$
\end{itemize}

\vspace{0.5in}

We now discuss these in turn, distinguishing between those demonstrated by
\COBE\ alone, and those demonstrated by the measurements at smaller angular
scales that have been made since \COBE.

\vspace{0.2in}
\noindent{\bf 2.A.1 Gravitational instability}
\vspace{4pt}

\noindent
Perhaps the most useful result of the \COBE\ anisotropy data is the
normalization of models of structure formation at large-angles, where
the fluctuations in the matter and photons are expected to be in the
linear regime.  In today's
favoured models of structure formation these large-angle anisotropies
directly measure the amplitude of the gravitational potential on very large
scales, allowing a theoretically clean and precise normalization of the
matter power spectrum.  This normalization, it turns out (e.g.~\cite{BEW}),
is in the right ball-park to explain the amplitude of galaxy
clustering (and with a little tuning of this or that parameter it is easy to
get complete consistency).
This is a vindication of our ideas that galaxies grew gradually under the
action of gravitational instability.

Before the \COBE\ anisotropy was announced it was often claimed
(e.g.~\cite{EU}) that extra physics would be needed if the results turned out
to yield yet more upper limits; right up to the DMR announcement it was also
commonly perceived that inflationary adiabatic models had difficulty having a
high enough amplitude to form structure without violating CMB limits
(e.g.~\cite{textures}).
The fact that the anisotropies were measured at the levels predicted, in
models with cold dark matter and adiabatic fluctuations, showed that there
is no need to invoke extra magical processes to form structure by the present
day.  However, since the photons prevent baryonic matter from collapsing
before recombination, we infer that the gravitational potentials had to be
dominated by matter which was not prevented from collapsing by photon
pressure, i.e.~matter that was not coupled to photons and was `dark'.
The realization, from studies of the galaxy distribution in the local Universe,
that matter formed `bottom up' rather than `top down' constrains the velocity
dispersion of the dominant dark matter component to be extremely small --
the dark matter must be mostly {\it cold}.

\vspace{0.2in}
\noindent{\bf 2.A.2 Recombination}
\vspace{4pt}

\noindent
Here we are moving beyond simply an interpretation of the \COBE\ data,
and looking at the large number of detections of anisotropy at degree
and sub-degree scales (see Fig.~1).
Early reionization of the Universe gives increased optical depth to Thomson
scattering from the present back to the epoch of reionization.
The extreme
case is a universe which did not (re)combine at all and remained ionized
for all time.
Multiple scattering erases existing anisotropies on scales smaller the horizon.
Thus reionization leads to damping of primordial anisotropies on small scales
(\cite{SugSilVit,Reion}).

The presence of fluctuations at $\ell\gtrsim100$ is clear evidence that the
Universe was not reionized at a very early epoch.  We can be confident that the
Universe recombined at $z\simeq10^3$, then remained largely neutral until some
redshift $z_{\rm reion}$, after which it was largely ionized (as implied by
the absence of Gunn-Peterson absorption in the spectra of high-$z$ quasars).
The precise value of $z_{\rm reion}$ derived from fits to the data
depends on the cosmological model, but is typically
$z_{\rm reion}<50$ (\cite{SSW,Teg}).

\vspace{0.2in}
\noindent{\bf 2.A.3 Degree scale power}
\vspace{4pt}

\noindent
We believe that Fig.~1 shows a peak in power in the anisotropies at scales
around a degree.  The {\it precise\/} position of this peak, how high it might
be, and whether it contains any substructure, are not so clear
(see e.g.~\cite{SSW,Hancock,Lineweaver,Bartlett,BonJafKno,Teg}).
However, it is striking that this feature is in the general location of the
main acoustic peak predicted by currently favoured models, based on the
angular size of the horizon at last scattering.
It is worth stressing that this prediction was made more than a decade before
the experiments were performed (see for example \cite{DZS}).
We expect the location of the peak to be determined definitively quite soon,
by upcoming ground based and balloon experiments, interferometers and
\MAP, leading to very strong observational constraints on the angular
diameter distance back to last scattering ($z\sim 10^3$).

\vspace{0.2in}
\noindent{\bf 2.A.4 Polarization}
\vspace{4pt}

\noindent
It is a fundamental prediction of the gravitational instability paradigm
that the CMB anisotropy is linearly polarized.  In inflationary CDM-like
models the level of polarization is a few percent of the anisotropy, and
thus extremely small in absolute terms.
There are already many limits on the polarization of CMB anisotropy
(see \cite{Polar} for a list), however they are all nearly an order of
magnitude larger than the theoretical predictions.
The fact that the CMB is not `very' polarized tells us important information
about the conditions at the last scattering epoch.  That the CMB is not very
circularly polarized, for example, indicates that there were no large magnetic
fields present at last scattering (see also \S2.B.4), although we are only
aware of very stringent upper limits at the smallest angular scales
(\cite{Paretal}).

\vspace{0.2in}
\noindent{\bf 2.B.1 Inflation}
\vspace{4pt}

\noindent
We put this item at the very top of our `B'-list since we feel the weight of
evidence is becoming very strong for something akin to inflation
(for a discussion of whether inflation is really a testable theory,
see \cite{BarLid}).
To avoid semantic arguments, it is important at the outset to be clear
about the meaning of `inflation'.
Here we refer to a period of accelerated expansion in the early Universe.
This is the only known mechanism for making an isotropic and homogeneous
universe, and at the same time generates apparently acausal adiabatic
fluctuations, i.e.~fluctuations in spatial curvature on scales larger than
the Hubble-length at a particular epoch.
We do not intend `inflation' to carry the additional baggage of an inflaton
field with a well-defined potential, connected with particle physics, etc.,
although ultimately we would all like to see the mechanism of inflation find
a realization in a well motivated theory of fundamental physics.

The amplitude and power spectrum of CMB anisotropies from degree-scales
up to the largest scales probed by \COBE\ seem to indicate that super-horizon
size adiabatic fluctuations exist.
Our first hint comes from the normalization of the large-scale anisotropies
relative to the matter (see e.g.~discussion in \cite{Comment}).
On dimensional grounds we expect that the amplitude of the temperature
fluctuations be ${\cal O}(\Phi)$ where $\Phi$ is the large-scale gravitational
potential.
In adiabatic models a cancellation (\cite{WhiHu}) between intrinsic
anisotropies and gravitational redshifts means that the coefficient is
reduced to $1/3$, i.e.~$\Delta T/T=-\Phi/3$ (\cite{SacWol}).
In the simplest isocurvature models the coefficient is $2$.
Since, as we mentioned before, our currently popular theories `work', there
is little room to absorb a factor of 6 in relative normalization.
Of course this alone is not proof of adiabatic fluctuations.

Our next piece of observational evidence is the angular scale of the `peak'
in power.  The structure of the peaks (locations, separations, relative
heights) is a strong discriminator between adiabatic and isocurvature models
(\cite{HuWhi}).  In almost all isocurvature models the peak is shifted to
smaller angular scales.
Since we observe excess power at about the right place for adiabatic
fluctuations in a flat universe, there is little room for either spatial
curvature or isocurvature fluctuations (and the combination is particularly
disfavoured!).
Since the current evidence for a peak, in contrast to a rise, is modest
we have put this in our `B'-list.  The observational situation is likely to
change rapidly.
In the future we can hope that detection of polarization on degree scales
will finally pin down the fluctuation type beyond any argument
(\cite{HSW,Polar}), but this is a difficult measurement due to the low levels
of signal.

Thus there is reasonable evidence
for adiabatic fluctuations in a spatially flat
universe.  The latter has long been hailed as a `prediction' of inflation.
The former is also tantamount to a `proof' of inflation, in the sense that
the only {\it causal\/} means for generating nearly scale-invariant adiabatic
fluctuations is a period when ${\ddot a}>0$ in the early Universe
(see e.g.~\cite{HuTurWei,Liddle}).
Of course this condition is neither entirely necessary nor sufficient.
On the sufficiency side, it is no doubt possible to imagine inflationary
models which have fluctuations of an entirely different character, but it
would seem pathological to deliberately avoid explaining density perturbations.
And on the necessary side, one could in principle imagine some early Universe
physics which somehow mimics the effects of inflation by producing
super-horizon adiabatic modes, and yet is not inflation.
We would argue that this is a purely semantic distinction:
if it looks like inflation and smells like inflation, then let's
call it inflation while leaving open the possibility that current
inflationary ideas may one day be shown to be part of some better paradigm.
In the same vein it may also be argued that some Planck-era physics somehow
generates apparently acausal modes.  Again we would say that is either
isomorphic with inflation, or simply an attempt to push the question of
initial conditions into the realm of metaphysics.

\vspace{0.2in}
\noindent{\bf 2.B.2 Space-time structure}
\vspace{4pt}

\noindent
We have already mentioned that the extreme isotropy of the CMB is a strong
indication that the FRW metric is an excellent approximation to the large-scale
properties of space-time.
Strong quantitative limits on the rotation and shear of space-time for
specific Bianchi models have been obtained from the \COBE\ data
(\cite{BunFerSil,KogHinBan}).  And limits on the geometry for general models
can be placed at the $\sim10^{-5}$ level (\cite{George}).

CMB anisotropies probe the Universe on the largest accessible
scales, and so they also constrain things like the large-scale topology.
There are quite stringent constraints in the simplest background models
(\cite{SteScoSil,deOSmoSta}).
However, in principle there may yet be observational consequences
for compact topologies, in an open universe in particular
(\cite{LevBBS,CorSpeSta,SouPogBon}).
Exactly how stringent the current constraints are,
for general classes of cosmology on the largest scales, is still a matter of
debate.  Nevertheless,
we probably know at this point that the Universe isn't {\it very\/}
strange on Gpc scales, quite an advance over our previous ignorance.

\vspace{0.2in}
\noindent{\bf 2.B.3 Gravity waves}
\vspace{4pt}

\noindent
If whatever produces the initial density perturbations doesn't discriminate
on the basis of perturbation type we would expect that scalar, vector and
tensor fluctuations would be produced at early times in roughly equal amounts.
The vector modes, representing fluid vorticity, decay with time and so would
not be present after a few expansion times.  Thus we would expect today to
see only scalar (density) perturbations and tensor (gravity wave)
perturbations.
Both of these types of perturbation would give rise to large-angle
anisotropies, though only the former will seed large-scale structure.
Due to the aforementioned close consistency between the amplitude of the
clustering on galaxy scales and the anisotropy seen by \COBE\ there is a
limit to how much the gravity wave signal can contribute to \COBE.
Roughly speaking, the tensor to scalar ratio $T/S<1$
(see \cite{Salopek,MarSta,ZibScoWhi}).
If the tensor perturbations are not too different from scale-invariant
this means that the possibility of seeing primordial gravity waves with
detectors such as LIGO or LISA is small
(\cite{KraWhi,Turner,Lid,CalKamWad}).

As has been argued by Lyth (\cite{Lyth}), the low-level of gravity waves is
good news for our current ideas about realizing inflation in simple particle
physics inspired models.  In the most popular models today, the scalar
modes are expected to dominate over the tensor modes by many orders of
magnitude.  The expectation is therefore that the tensor signal may not be
measurable with any existing or planned experiments, or conversely that a
positive detection of gravity waves would have profound implications for our
ideas about inflation.  However, for the time being, the constraints on the
gravity wave contribution have not reached the level where we learn
much about early Universe physics -- that will await future experiments.

\vspace{0.2in}
\noindent{\bf 2.B.4 Physics at $z\sim10^3$}
\vspace{4pt}

\noindent
It is possible to use the fact that the CMB anisotropies are largely as
expected to limit the magnitude of any surprises at the last-scattering
epoch.  The arguments are much akin to those using the observed abundances
of the light elements and Big Bang Nucleosynthesis theory to limit `exotic'
physics at early times.  If something `exotic' would dramatically alter
the theoretical predictions, it can be strongly constrained.
A great many possible physical effects have been studied, but here we will
list only a few things for which it is already possible to place observational
bounds.  Strong limits exist on domains of anti-matter (\cite{KinKolTur}),
particle decays near $z\sim10^3$ (\cite{PieBon}), primordial voids from an
early phase transition (\cite{SakSugYok}) and primordial magnetic fields
(\cite{Barrow,SubBar}), among other things.

\section{The Future}

\noindent
It was stated in the early 1960s, shortly before the discovery of the CMB,
that there were only 2\onehalf\ facts in cosmology
(by Peter Scheuer, see \cite{Longair}).
In a similar spirit, we have argued that there are perhaps 4 facts and 4
half facts currently known from CMB anisotropies.

It has been recognized for some time that these anisotropies may answer some
of our most fundamental questions about the Universe.
The current CMB data already indicate that gravitational instability, in a
mostly cold dark matter dominated universe, amplified initially small adiabatic
fluctuations into the large-scale structure that we see today.
There is the potential to show what inflationary-like process happened in the
early Universe.  And ultimately, the precise shape of the angular power
spectrum holds the key to determining many of the fundamental cosmological
parameters, either directly or in combination with other measurements.

\begin{figure}[t]
\begin{center}
\leavevmode
\epsfxsize=16cm \epsfbox{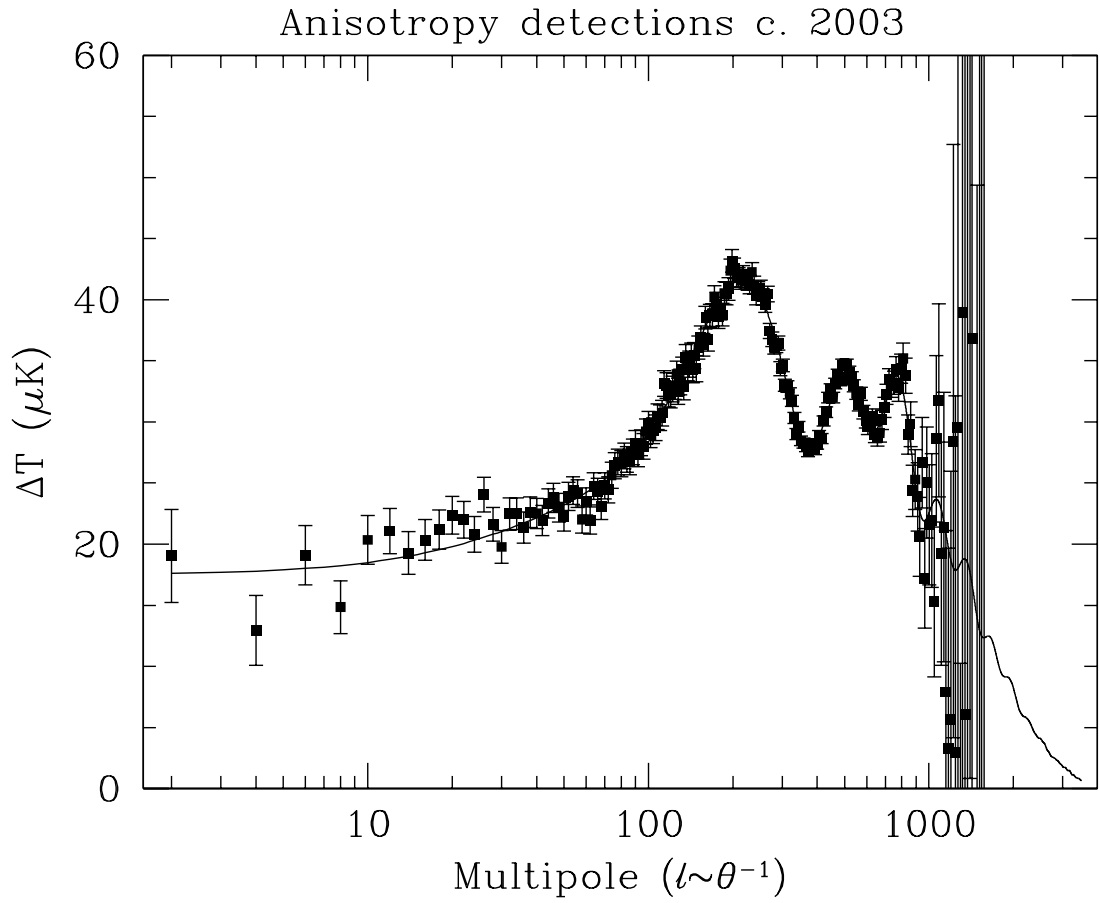}
\label{fig:future}
\caption{The future of CMB anisotropies as possibly detected by
\MAP\ and by Planck, representing the potential state of knowledge
roughly 5 and 10 years after the present.}
\end{center}
\end{figure}

\placefigure{fig:future}

However, while it is interesting to track progress in this field and to
speculate on what it all means, it seems clear that
theorists have had long enough to manoeuvre
that the present data no longer strongly constrain any popular cosmological
model.
With the coming of long duration balloon flights, the imminent launch of the
\MAP\ satellite, and the commissioning of three new CMB interferometers,
we expect that to change.
The BOOMERANG team has already had a successful long duration balloon
flight, and the analysis of that data set is eagerly awaited.  Similar
flights will undoubtedly follow, along with other large data sets from
new ground-based experiments.  The race is on, since \MAP\ is scheduled
for late 2000.
A little later, sometime around 2006, will see the launch of the Planck
Surveyor.  Planck should supply us with essentially cosmic-variance limited
information on all the angular scales relevant to primary anisotropies, over
the full range of relevant frequencies.
Figure~2 is an estimate of how well the power spectrum might be
constrained after \MAP\ and after Planck.
With the proliferation of high precision data
future `A' and `B' lists will be correspondingly longer and more
detailed.  Our attempt at
prognostication is represented in our list `C':

\vspace{0.5in}
\begin{itemize}
\item[{\bf C1}]\ Cosmological parameters will be precisely determined
\item[{\bf C2}]\ Polarization will be measured over a range of scales
\item[{\bf C3}]\ We will learn about early Universe physics
\item[{\bf C4}]\ We will learn much about non-linear astrophysics
\end{itemize}
\vspace{0.5in}

Item~1 is in fact already happening, as discussed earlier.  However, the
current set of anisotropy data is not very constraining, since there is
enough parameter freedom to fit models with quite different values of any
individual parameter (\cite{Teg}).  This situation
will undoubtedly improve in the future (unless of course
{\it none\/} of the current models fits the data, which is surely the most
exciting prospect of all!).
Certainly some degenerate parameter combinations will continue to exist in
the model space (particularly in models with the same `angular diameter
distance'), but these degeneracies can be broken through combinations
with other astrophysical data sets (\cite{White,EHT}).
If systematic errors can be kept under control, the combination of Planck
and data from redshift surveys will be particularly powerful at determining
the cosmology.

Item~2 will be difficult, but we have no doubt that it will happen.
\MAP\ may yield some information, how much is difficult to estimate without
more insight into the foregrounds.  Currently planned ground-based
experiments may also give detections.  And Planck should provide polarization
measurements over a reasonable range of scales.
However, a full investigation of CMB polarization (and certainly the `curl'
or $B$-mode component produced by tensors) may have to await an experiment
even beyond Planck.

Item~3 potentially involves information from both 1 and 2.  Ultimately we
will learn something about high energy physics through understanding the
way in which fluctuations were laid down in the early Universe, whether
this involves discriminating any tensor component, measuring a changing
spectral index, non-Gaussian signatures, or something else.
Since the relevant energies are so far beyond what is achievable in
particle accelerators, it is likely that cosmological phenomena will be the
only way of constraining such models for quite some time.
In addition to the `initial conditions', the evolution of the fluctuations
will provide us with information on the properties of the dark matter in the
Universe which may tie in directly to particle physics theories at the
electroweak scale.

Item~4 includes a whole suite of potentially measurable effects, which can
be thought of as processing the primary anisotropies.  Examples include
gravitational lensing, non-linear potential growth, Sunyaev-Zel'dovich
effects, details of the reionization process, and extragalactic sources.
There is a grey area between what is considered cosmic signal and what is
considered a `foreground'.  But whatever you call it, there is little
doubt that data from the Planck mission, for example, are likely to be mined
for many years for the additional astrophysical information they contain.

We expect rapid experimental progress in the next few years, and we trust
that theoretical effort will be similarly feverish (\cite{Bond}).
As a result, there will no doubt be more physical processes uncovered which
affect CMB anisotropies.
At present the CDM-dominated inflationary paradigm looks like it's in
pretty good shape.
Our `C' list may end up being quite inaccurate, and we can even imagine
trouble for some entries in our `B' list.
However, the spectral information from the CMB, together with the `A' list,
provides a very solid foundation for the physics which
generates the anisotropies.
Therefore we are confident that whatever proves to be the ultimate such list,
a thorough investigation of CMB anisotropies will hold the key to learning
about the background space-time and formation of large-scale structure in
the Universe.

\acknowledgments
DS is supported by the Natural Sciences and Engineering Research Council of
Canada.  MW is supported by the NSF.

\newpage

\newpage
 
\newpage
 
\begin{figure}[t]
\begin{center}
\epsfxsize=16cm \epsfbox{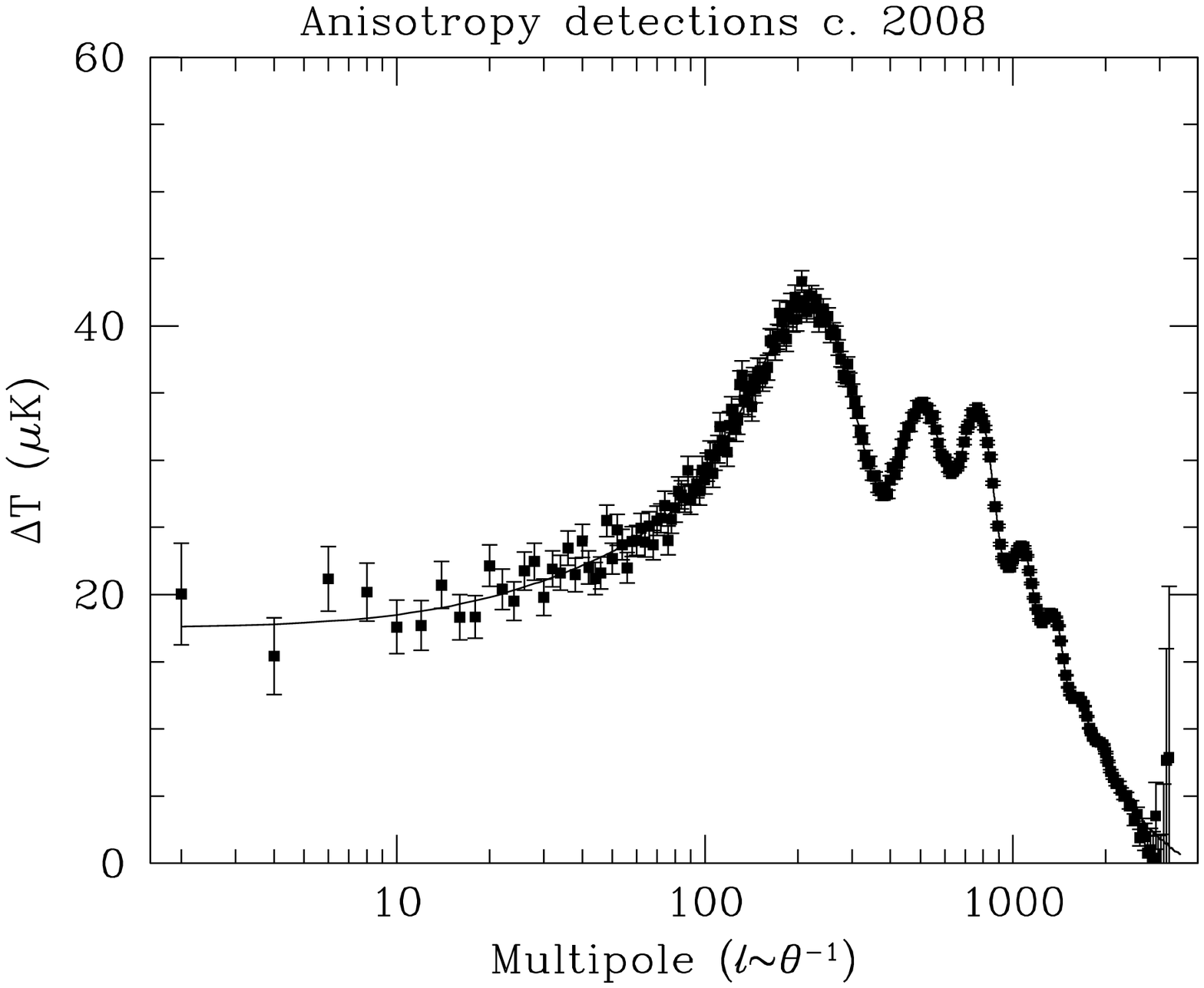}
\end{center}
\end{figure}

\end{document}